\documentclass[12pt,preprint]{aastex}
\shorttitle{Fluorescent H$_2$ Emission Near T Tau}
\begin{document}

\title
{THE SPATIAL DISTRIBUTION OF FLUORESCENT H$_2$ EMISSION NEAR T TAU}

\author{Jos\'e Saucedo \altaffilmark{1,2}, Nuria Calvet \altaffilmark{1}, 
Lee Hartmann 
\altaffilmark{1}, and John Raymond \altaffilmark{1}}

\altaffiltext{1}{Harvard-Smithsonian Center for Astrophysics, 60 Garden 
St.,
Cambridge, MA 02138, USA, \email{\scriptsize jsaucedo@cfa.harvard.edu, 
ncalvet@cfa.harvard.edu, hartmann@cfa.harvard.edu, 
raymond@cfa.harvard.edu}}

\altaffiltext{2}{Instituto de Astronom\'{\i}a,
UNAM, Ap. Postal 70-264, Cd. Universitaria, 04510 M\'exico D.F.,
M\'exico}

\begin{abstract}

New subarcsecond FUV observations of T Tau with HST/STIS show spatially 
resolved structures in the 2"$\times$2" area around the star. The 
structures show in multiline emission of fluorescent 
H$_2$ pumped by Lyman $\alpha$. One emission structure follows the 
cavity walls 
observed around T Tau N in scattered light in the optical. 
A temperature
of $\ge$ 1000K is required to have enough population in the
H$_2$ to produce the observed fluorescent lines; in the cool environment
of the T Tau system, shock heating is required to achieve
this temperature at distances of a few tens of AU.
Fluorescent H$_2$ along the cavity wall
represents the best evidence to date  for the action of 
low-density, wide-opening-angle outflows driving
cavities into the molecular medium
at scales $\le 100$ AU.
A southern region of emission consists of
two arcs, with shape and orientation
similar to the arcs of H$_2$ 2.12 {\rm $\mu m$}
and forbidden line emission crossing
the outflow associated with the embedded system T Tau S.
This region is located
near the centroid of forbidden
line emission at the blueshifted lobe of the
 N-S outflow.

\end{abstract}


\newpage

\section{ Introduction}
T Tau is a multiple system of pre-main sequence stars, composed of
an optically-visible K0 star, T Tau N, and a heavily-extincted 
system, T Tau S \citep{D82}, which is itself binary \citep{K00, Kh00, D02}.
 The T Tau system is surrounded by
an infalling envelope of a few thousand AU \citep{C94}, which is in the
process of being disrupted by powerful outflows \citep{M96}.
Two bipolar outflows have been identified in the system
\citep{B88, B94}. One of the outflows runs 
NW-SE
and has been associated with the embedded pair T Tau S.
Analysis of the forbidden line emission indicates that
this outflow
is poorly collimated \citep{B94, S99}. Bright 
arcs
of forbidden line emission \citep{R95} and near
infrared H$_2$ line emission \citep{H96, H97}
are found crossing this outflow at scales of $\sim$ 2" - 14"  
($\sim$ 280 - 2000 AU at the distance of Taurus,
140 pc, Kenyon et al. 1994). 
Burnham nebula, $\sim$ 8" south of the
system, is associated with the blueshifted lobe of this outflow;
it continues in 
jets detected out to $\sim$ 0.7 pc \citep{R97}.
The other outflow runs  E-W and is associated
with T Tau N. It is less bright in the optical,
although high velocity forbidden line emission
is detected towards the west, ending in Herbig
Haro object HH155 located at
the edge of the reflection nebulosity NGC 1555 (Hind's
nebula), $\sim$ 35" west from the star. $^{12}$CO emission associated with the 
E-W
outflow shows blue and redshifted lobes of moderately
high velocity (Schuster et al. 1994), surrounded
by rings of $^{13}$CO emission at lower velocity;
these rings have been interpreted as arising at the wall of
the cavity  in the envelope opened by the E-W outflow
(Momose et al. 1996).

The structure of the outflows and molecular gas and dust 
very near T Tau (inner 2"x2")  is  
uncertain.  
Observations in the V, R, and I filters with the
WFPC2 camera on the Hubble Space Telescope
($HST$) by Stapelfeldt  (1998, S98) showed a scattered light 
structure
whose morphology suggests an illuminated 
outflow cavity 
in the circumsystem envelope.  Spectroscopic mapping in the K (2.01-2.42 
{\rm 
$\mu$}m) 
and H (1.5-1.8 {\rm $\mu$}m) bands by \cite{K01} probed the inner
2"$\times$2" of the T Tau system; unresolved Br$\gamma$ emission at
the positions of T Tau N and S was observed, but no infrared H$_2$ emission 
was 
detected.

In contrast, space-based observations in the UV of molecular hydrogen have 
indicated
structure on small scales near T Tau.
\cite{B81} detected extended H$_2$ fluorescent emission 
in T Tau N and its adjacent nebula in a low resolution large
aperture spectrum from the
IUE satellite.  
\cite{V00} refined the
location of the H$_2$ emission by comparing data from the Goddard High
Resolution Spectrograph (GHRS) on $HST$ in G140L mode with all the 
available 
IUE 
spectra of T Tau.  The H$_2$ line fluxes were found to be
larger in the IUE spectra 
suggesting that the molecular hydrogen emission
extended beyond 0.2" from the star, the size of the Small Science 
Aperture in GHRS.  
In this paper,  we present new data from 
the Space Telescope Imaging Spectrograph
(STIS) on board $HST$ which
illustrate the spatial distribution of fluorescent H$_2$ emission in the T 
Tau
system, with important implications for the interaction of outflows and 
circumstellar
matter.

\section{Observations}

T Tauri was observed with $HST$ on 2000 February 7 in
program GO8317, using the STIS MAMA detector with the G140L 
grating. A 2" long slit with a PA=$30^{\circ}$ was employed, and the
data were taken in binned pixels mode with an exposure time of 1622 s.  
The spectral range covered from 1118 to 1715 {\AA} with a 
dispersion of 0.58
{\AA} per pixel resulting in a
 resolution of $\sim $1
 {\AA} at 1500 {\AA}, and a plate scale of 0.024" per pixel.  
Standard CALSTIS pipeline procedures were used to reduce the data.

Figure 1 shows the position-dispersion image.  Faint emission
is seen extending roughly 1.6" $\sim$ 240 AU southward from the ultraviolet 
image of T Tau 
N.
Some extended emission is seen to the north, but to a much smaller extent.
The spectra of the extended emission were extracted using standard 
IRAF/STSDAS/X1D tasks,
although the extraction was difficult  because the spectra have low 
signal-to-noise 
 ($\sim$ 8 for the strongest lines) and because extended structures fill the
large aperture in the dispersion
direction. For the extraction,
the position of the background was taken outside of the range
corresponding to the geocoronal
Lyman    
$\alpha$ and O I lines and far from the extended
emission; the background was the same for all the spectra, averaged over 10 
pixels.
The signal along the extended emission is weak, roughly 20 times
weaker
than the stellar spectrum. Line emission accounts for about
50\%
of the total signal.
We attempt to identify the strongest
emission features, recognizing that there may be weaker, more diffuse
emission that we cannot trace.

Figure 2 shows the spectrum of T Tau N, scaled by a factor of 0.2, and
spectra extracted in 10 pixel = 0.24" boxes at various offsets from 
the star along the slit orientation.
Lines from
the highly-ionized species Si IV 1393/1402, C IV 1458/1550, and He II 1640 
are present in the
stellar spectrum; the corresponding emission features 
in the spectra offset at $\pm$ 0.39" are consistent with the expected
wings of the instrumental point spread function for the stellar emission.
The other line features in the extended nebulosity
are consistent with fluorescent H$_2$ emission.
As shown in Figure 2, 
the strongest lines arising from
de-excitation
of the rotational-vibrational level (v',J'=1,4) of the electronic state
$^1\Sigma_u^+$ populated by the (1-2)P(5) transition can be identified.
There is also a suggestion that the strongest lines of the de-excitation of
the (v',J'=1,7) level, populated by the (1-2)R(6) transition are present.

Table 1 shows the observed fluxes of the fluorescent
lines due to the (1-2)P(5) transition in several offset
spectra. The fluxes are
measured above 
the estimated continuum.
The last column shows the ratios of the line fluxes at $\lambda 1446$ and 
$\lambda 1504$ \AA, 
 where the fluxes
have been corrected for reddening
with A$_V$=2.2 using the HD29647 
extinction law \citep{C03}.
The ratios are consistent with the optically thin ratio (0.73),
within the uncertainties given by their low S/N.

\section{Spatial distribution} 
\label{espacial}

To make the line identifications shown in Figure 2,
it is necessary to impose wavelength shifts to
the spectra of the extended emission.
If these shifts were interpreted as velocities, they would require motions
of $\sim$ 2000 {\rm km s$^{-1}$}  or more,
much larger than the velocities previously reported in the region
($\sim$ 45 {\rm km s$^{-1}$}; 
\citep{B94, E98}.
 A more likely explanation for the wavelength shifts
is that they are due to spatial shifts of the emission regions
relative to the star within the large (2") aperture.
In this case, the wavelength shift $\Delta \lambda$
would imply a
displacement $ x = s \Delta \lambda / m $ from the axis
of the slit,
where $s$ is the plate scale and $m$ is the dispersion. 
All the offsets are measured from the position
of T Tau N.

Figure 2 indicates at the right the offsets $x$ required
to fit the H$_2$ lines
at each $y$ offset from the star.
Two spatially-distinct regions of H$_2$ emission can be identified.
One region extending to $y \sim 0.7$" southwest from the star,
which is shifted slightly westwards,
and a region covering
$0.7" < y < 1.6"$ from the star, which has
an eastward displacement.
These regions can be discerned in Figure 1. 
 The northern
region, labeled by N in the sketch below the spectrum in
Figure 1, can be seen as three features
extending to a distance of about 0.7" southward
from the star, at PA $\sim 210^{\circ}$,
corresponding to the H$_2$ line pairs at
1490/1505 {\AA}, 1431/1446 {\AA}, and
1547/1562 {\AA}, although in the latter case
the structure near the star is difficult to see due to the strong stellar
C IV 1550 {\AA} emission.
Further south, these emission features
seem to disappear; however, similar pairs of emission lines
can be seen shifted roughly 25 {\AA} blueward, extending nearly
vertically in the image to about 1.6" south.  This 
region of emission is labeled S in  Figure 1. 
The N region extends to the north of the star as well,
but it is difficult to identify
spectral features in it.
After taking into account the position angle of the slit,
we find that the spatial location of the upper N emission region 
appears to correspond to the
the brightest part of the southern optical scattered light structure imaged 
by S98, 
while the lower emission region roughly follows the outer border of the
scattered light nebula, although displaced from it.

To obtain more information about the spatial distribution
of the emission, we constructed a model and
compare the predicted
emission with the observed spectrum. If the spatial
distribution of the emission 
at a given offset $y$ is given by {\it w(x,y)}, and the intrinsic
emission at each point is $G_{\lambda}$, then the
observed flux at wavelength $\lambda$ is given by
the convolution of {\it w(x,y)} and  $G_\lambda$,
\begin{equation}
{\it F}_{\lambda}(y)=a \sum_{i=1}^{N}  G_{\lambda-\Delta 
\lambda_i}w(x(i),y)
\end{equation}
where $a$ is a scaling
constant,
{\it x(i)} is the spatial shift across the slit from its axis,
$\Delta \lambda_i$ is the wavelength shift corresponding to
$x(i)$, and $N$ is the number of pixels across the slit.

We used as template $G_{\lambda}$ the spectrum of 
HH 43, a low-excitation Herbig-Haro object,
which clearly shows the fluorescent lines we have identified
\citep{Sc83}.  The spectrum is an average of the three IUE large aperture
spectra with the highest exposure times, SWP31828, SWP23881, and SWP24924,
 taken from the Multimission Archive at the Space Telescope Science Institute,
which have been
processed under the NEWSIPS extraction pipeline.
The combined signal to noise of the resultant HH spectrum is
8-13.
Since we are interested in the relative distribution
of the H$_2$ intensity, we 
normalized $w(x,y)$ to 1 and scaled
$G_{\lambda}$ with the constant {\it a}
to fit the observed spectrum
of the extended emission at offset $y$ = -0.39".

The predicted spectra at several offsets $y$ from the star
are shown in Figure
3 compared to the observed spectra smoothed to the IUE resolution of 6 
{\AA}; 
the corresponding spatial map of the emission $w(x,y)$
is shown in Figure 4.  

The H$_2$ emission in the southern part of region N
(N1 in Figure 4)
appears to follow the brightness and spatial
distribution of the scattered light image 
of S98, with some indication that
the H$_2$ is somehow spatially more
concentrated than the optical region. The H$_2$ and
optical emission are similar from $\sim$ 0."45 to $\sim$ 0."65
from the star,  but around $\sim$ 0."45, the H$_2$ emission is $\sim$ 30\% 
narrower
than the optical scattered light.
Consistently, the H$_2$ emission in the northern part of
region N (N2 in Figure 4) is low, 
since the north part of the optical emission is two magnitudes fainter
than the
southern portion at the same distance from T Tau N (S98).

The resulting 
predicted spectra for region N1 are shown at -0.39" and -0.62" offsets in 
Figure 
3.
The spectral agreement is reasonable considering the low S/N
of the spectra and the contamination by stellar emission at  
-0.39".  The principal failing of the model is the lack of predicted
emission near 1600-1620 {\AA}, which is identified in the spectrum of T Tau 
N as 
a 
mix of CII an Fe II by Valenti et al. (2000).
Fe II emission may arise from the $\sim$40 $\rm km~s^{-1}$
J-shocks discussed below as sources of Lyman $\alpha$ photons
for the lower emission region.
It is also likely that the broad
Lyman $\alpha$ emission profile of T Tau produces some of the
fluorescent sequences observed in Mira \citep{W02}
but not in low excitation HH objects.  Some of these sequences
include emission near 1610 {\AA}.

For region S, $y \le -0.85"$, 
the observed line profiles are too narrow to
resolve the emitting regions;
we found the best fits by introducing two narrow
structures
at slightly different positions, labeled S1 and S2 in Figure 4.
The agreement between predicted and observed spectra in Figure 3
is reasonably satisfactory except
for a feature near 1600-1620 {\AA} in the $y = -0.62"$ offset spectrum.
We should also emphasize again that it would be difficult for us to detect 
very
spatially-extended H$_2$ emission in our large aperture data.

\section{Physical conditions}
\label{physical}

\subsection{Estimates of Lyman $\alpha$ luminosities}

The total observed flux in the
strongest H$_2$ line we observe,
(1-7)P(5), $\lambda$= 1504.8 \AA, can be written as
(cf. Jordan et al. 1978)
\begin{equation}
F_{H_2} =3\times 10^{-16} \, N(2,5) A_{H_2} \, \bar J_{\nu}\, {\rm \; 
erg\;cm^{-2}\;s^{-1}},
\label{flujo}
\end{equation}
where $N$(2,5) is the column
density in the vibrational-rotational level (v",J"=2,5) of H$_2$,
 $A_{H_2}$ is the H$_2$ emitting area projected in the sky
in arcsec$^2$, and
$\bar J_{\nu}$ is the mean
intensity of the Lyman~$\alpha$ radiation capable of exciting
the (1-2)P(5) transition\footnote{From now on, the
bar notation refers to any quantity $\bar f=\int\phi_\nu f_\nu d_\nu$, with
$\phi_\nu$ the line profile of the H$_2$ line, centered at
$\nu_0 + \Delta \nu =\nu_0 + \nu_0\Delta v/c$
from the Lyman $\alpha$ line center at $\nu_0$.
For the  (1-2)P(5) transition, $\Delta v = +98 {\rm km \, s^{-1}}$.}, in
${\rm \; erg\;cm^{-2}\;s^{-1} \; Hz^{-1}}$.
We have used the line parameters 
$B_{(2,5\to1,4)}=1.976\times 10^8$ erg$^{-1}$ cm$^2$ s$^{-1}$,
$A_{(1,4\to7,5)}=1.969\times 10^8$ s$^{-1}$ and
$\Sigma_j A_{(1,4\to j)}=1.7125\times 10^9$ s$^{-1}$ from \cite{A00}.

 If we assume that the H$_2$ emission
region N is irradiated by stellar Lyman $\alpha$
and that there is no extinction between the star and
the region, then
the mean intensity $\bar J_{\nu}$ reaching the region
is related to the specific
intensity $\bar I_{\nu}$ leaving the star by
\begin{equation}
\bar J_\nu \approx {1 \over 4\pi} { {\bar I_\nu A_{Ly \alpha}} \over  
d^{2}},
\end{equation}
where $A_{Ly \alpha}$ is the area of the Lyman $\alpha$ emitting region 
on the star
and $d$ the distance
between the star and the site where the H$_2$ molecules are
pumped. In turn, $\bar I_{\nu}$ can be expressed in terms of the 
flux $\bar F^{obs}_{Ly \alpha}$ of Lyman $\alpha$ observed at Earth,
as $\bar I_{\nu} =\bar  F^{obs}_{Ly \alpha} D^{2} / A_{Ly \alpha}$
(assuming isotropic emission)
where $D$ is the distance to the star.
Combining these expressions, we can write 
\begin{equation}
\bar J_\nu \approx 3.3 \, \times 10^9 { \bar F^{obs}_{Ly \alpha} \over 
\theta^{2}} {\rm 
erg 
\;cm^{-2}\;s^{-1}\;Hz^{-1}\;sr^{-1}},
\label{jotanu}
\end{equation}
 where $d/D$ 
has been approximated by the apparent angular separation
between the star and the H$_2$ emitting region
on the sky, $\theta$, in units of arcsec in eq.(\ref{jotanu}).  

Inserting in eq. (\ref{flujo})
 we obtain
\begin{equation}
F_{H_2}  \approx 10^{-6} 
{ A_{H_2} \over \theta^{2}}
N(2,5) \bar F^{obs}_{Ly \alpha} {\rm \; 
erg\;cm^{-2}\;s^{-1}}
\label{flujo1}
\end{equation}

 We estimate a lower limit for $N$(2,5)
by taking the column density of the outermost layers
of the cavity where radiation of Lyman $\alpha$ can penetrate,
to a depth where $\tau_\nu \sim 1$.
The actual column density producing the emission may be larger than this if
the emitting region is not seen face-on.
This may be indeed the case, given the low inclination
to the line of sight, $\sim 10^{\circ}$, of the E-W
outflow estimated by \cite{M96}.
The optical depth for a transition line from an upper level {\it j} to a 
lower 
level {\it i} can be written as $
\tau_\nu=N_i B_{ij} h\nu_{ij} \phi_\nu (1-exp(-h\nu_{ij}/kT))/ 
4\pi \sim N_i B_{ij} h\nu_{ij} \phi_\nu /4\pi$, for a temperature of a few 
thousand K,
with $\nu_{ij}$ the frequency of the transition.
We assume that the H$_2$ line is Gaussian in shape,
with a Doppler width
$\Delta \nu_D = \nu_{ij} \sqrt{ v_{th}^2 + v_{tur}^2} /c$.
The agreement with the spectrum of HH43 (\S \ref{espacial}) suggests
a low excitation situation, so we adopt a
temperature T$\sim$1000 K to determine the thermal velocity
$v_{th}$, and assume a turbulent velocity $v_{tur}\sim$10 km s$^{-1}$
as expected from an oblique shock interface (\S 5.1).
Assuming 
the line profile is given by $\phi_{\nu}\approx 1/\pi^{1/2}\Delta \nu_D$,  
we get a 
column density $N(2,5)\approx 7 \times10^{14}$ cm$^{-2}$
from the condition $\tau_\nu \sim 1$ for the (1-2)P(5) transition.

Let $Q$ be the ratio 
between $\bar F^{obs}_{Ly \alpha}$ and the total observed flux $F_{Ly\alpha}$. 
 We cannot use the 
stellar Lyman $\alpha$ line profile in the spectrum
because it has been heavily absorbed by circumstellar
and interstellar extinction. So, we
estimate the value of $Q$ assuming 
that the Lyman $\alpha$ profile is similar
to that of Mg II 2800 {\AA},
since both 
are resonant lines with extended wings formed by partial redistribution.
 We use the high resolution line profile of Mg II k $\lambda$ 2796.3 in T 
Tau N,
obtained with STIS/NUV-MAMA and echelle grating E230M 
in our $HST$ program GO8627
(Fig. 5, Saucedo et al., in preparation).
At $\Delta v$= 98 km s$^{-1}$ 
from the center of the Mg II 2796.3 {\AA}
line, we determine $ Q \sim 1.4 \times 10^{-12}$ Hz$^{-1}$.
As illustration, we show in Figure 5 the H$_2$ line profile $\phi_{\nu}$
at $\Delta v$= 98 km s$^{-1}$ used for the calculation
of $\bar F^{obs}_{Ly \alpha}$.

The emitting area $A_{H_2}$ can be estimated from the height of the
extraction box used to obtain the spectra and the width of the relative 
intensity
map {\it w(x,y)} as $\sim$ 0.2$\times$0.24 arcsec$^2$. 
Taking the distance $\theta$ between the star and the H$_2$ region as 
$\sim$ 
0.5", 
an average flux of H$_2$ 1504.8 {\AA} $\sim 2.5 \times
10^{-15}$ ${\rm erg\; cm^{-2}\;s^{-1}}$ in region N (cf. Table 1), 
the estimated values of $N(2,5)$ and $Q$, and eq. (\ref{flujo1}),
we can write for the luminosity of the stellar Lyman $\alpha$ line
\begin{equation}
8 \times 10^{-3} L_\odot\leq 
L_{Ly \alpha} \leq  2.1 L_\odot
\label{lyalimits}
\end{equation}
where the limits arise from the uncertainty
in the reddening correction
 appropiate to the H$_2$ lines, since the
extinction along the line of sight of the nebulosity may not be the same as 
the
extinction of the star;
the lower limit comes from taking the observed
H$_2$ flux and the upper limit from dereddening this flux
by the extinction to T Tau N,
 $A_v=2.2$ \citep{C03}.
In addition, as discussed above, the actual value of $N(2,5)$
is likely to be higher because of projection effects,
making the required  Lyman $\alpha$ luminosity lower than
estimates in eq.(\ref{lyalimits}).
These limits for the Lyman $\alpha$ luminosity
are consistent with those obtained by scaling
by a factor of 30 the dereddened flux of the CIV
1548 line in our stellar spectra, as suggested by Ardila et al. (2002).
With F$_{\rm CIV }\sim 9.2 \times 10^{-12}$ $\rm
erg\; cm^{-2}
\;s^{-1}$, the scaled Lyman
$\alpha$ luminosity would be $\sim$ 0.35 L$_\odot$.

The estimated maximum Lyman $\alpha$ luminosity (eq. [\ref{lyalimits}])
is $\le$ 60 \%
of the
accretion luminosity of T Tau N 
 estimated from the 
NUV spectrum obtained simultaneously with the FUV spectrum,
${\rm L_{acc} = 3.2 L_\odot}$ \citep{C03}, 
so it can be accounted for by accretion energy.
We thus conclude that the  H$_2$ emission in 
region N is 
consistent with stellar  Lyman $\alpha$ pumping.

While the bright rims seen in region N are clearly illuminated
by T Tau N, 
the southern arcs 
S1 and S2  appear to be located (in projection) well inside
the envelope, roughly tracing the outer edge of the extended emission of
S98.
Lyman $\alpha$ from the visible star will be absorbed by dust in the
envelope so it cannot penetrate more than
one dust mean free path, $l\sim 3-4$ AU,
estimated from models of infalling envelopes \citep{W93} with typical envelope
parameters
from \cite{C94} ({$\rm \dot M=3\times10^{-6}\;M_\odot\;yr^{-1},\;
r_c=100\;AU,\;M=2\;M_\odot $}) and a dust opacity
at the  Lyman
$\alpha$ wavelength of 1538 cm$^2$ g$^{-1}$ \citep{O02}.
 It is therefore more likely that the required Lyman $\alpha$
emission comes from the shocks themselves,
as in the case of low excitation HH objects \citep{C95}.

If Lyman $\alpha$ 
is locally produced, then $\bar J_\nu
=\bar I_\nu /2$, so
we can write
\begin{equation}
L_{H_2} = 6.8 \times 10^{-6} N(2,5) Q L_{Ly \alpha}
\end{equation}
where $L_{H_2} = 4 \pi D^2 F$ is the H$_2$ luminosity.
Using similar values of $N(2,5)$ and $Q$ as in region N,
we get that the luminosity in Lyman $\alpha$ 
required to excite the H$_2$ fluorescence in region S is 
\begin{equation}
3.4 \times 10^{-4} L_\odot\leq   
L_{Ly \alpha} \leq  0.1 L_\odot, 
\end{equation}
 where again, the lower limit is given by the observed values of the H$_2$ 
flux, 
and 
the upper limit by assuming that the extinction is the same as towards T 
Tau N, 
although 
there is no argument to support the assumption of an homogeneous extinction 
in 
the 
whole region. 
 These limits are consistent with values of
the Lyman $\alpha$ luminosity in low excitation HH objects;
for instance, \cite{C95} find $ L_{Ly \alpha} \sim 0.13 L_\odot$
in HH 47A.

\subsection{Estimates of temperature and Hydrogen column density}

Fluorescence
requires a significant population in excited levels which also can lead
to infrared H$_2$ emission lines.  Thus, the observed upper limits 
to the infrared emission line fluxes 
in observations covering region N \citep{K01} can provide an 
additional constraint
on molecular column densities. 

An upper limit for the  2.12 $\mu$m H$_2$ line
can be estimated from the equivalent width
of 0.15 \AA, which is the detection threshold for features 
in the \cite{K01} observations. Using the continuum flux
of $\sim 3.1 \times 10^{-9} {\rm erg\;cm^{-2}\;s^{-1}\mu m^{-1}}$,
we obtain that the flux at the 2.12 $\mu$m H$_2$ line
needed for a 2 $\sigma$ detection would be
$9.3 \times 10^{-14}$ $\rm erg \;cm^{-2}\;s^{-1}$.

The flux at 2.12 $\mu$m H$_2$ in the optically thin limit is given by
$F_{2.12} = I_{2.12} A_{H_2} / D^2 = N(1,3) A_{31} h\nu_{31}/4 \pi 
A_{H_2}/D^2$,
where $N(1,3)$ is the column density of the upper
level of the transition. With the Einstein value $A_{31}=3.66\times 
10^{-7}s^{-1}$
\citep{G76} and the upper limit for the flux,
we obtain $N(1,3)\leq 2.24\times 10^{18}{\rm\;
cm^{-2}}$.

As mentioned before, we have only a lower limit to the actual N(2,5). 
With this lower limit and the upper limit for N(1,3),
we can obtain a lower limit to the excitation
temperature in region N through the Boltzmann relation
$ N(1,3)/N(2,5)
=1/1.57exp(-E_{(1,3\to 2,5)}/kT)$,
with $E_{(1,3\to 2,5)}/k \approx 6840$ \citep{H50}.
Using our estimates, we obtain 
$T \geq$
800 K, consistent with a low velocity shock.  From 
Figures 6 and 7 in \cite{J78}, 
the population of the levels with (v"J"=2,5) in this temperature range is 
$\sim$ 
12\% of
the populations with v"=2 and any J", and in turn, the population with 
v"=2 is 
$\sim$ 0.001\% of
the total population.  So, with $N(2,5) > $7$\times 10^{14}$ cm$^{-2}$, 
the column density of the H$_2$ molecule is
N(H$_2$)$ > 5.8 \times 10^{20}$ cm$^{-2}$.  

We have no measurements of H$_2$ 2.12 
$\mu$m in region S
to help constrain the 
temperature and total H$_2$ density.
Information on regions S1 and S2 can be obtained  from the [O I], [N 
II] 
and [S II] PV 
diagrams in \cite{B94}. From their Figure 5, 
regions S1 and S2 are located within 
one of the
multiple components identified in their work, component D,
which corresponds to the blueshifted lobe of the N-S outflow.
The line 
ratios [O I]/[S II], 
[N II]/[O I] and [N II]/[S II] around the location
of region S are 1.25, 0.35 and 0.28, respectively. 
Comparison with shock model predictions \cite{H87}
indicate a shock of velocity v=40 km s$^{-1}$.
For this shock velocity, the 
ratio of Lyman $\alpha$/[O~I] can be obtained from these models, yielding a 
value of $\sim$ 73; so, a rough estimate of the [O I] luminosity is $ 4.6 
\times 
10^{-6} L_\odot\leq   L_{[O I]} \leq  3.5\times 10^{-4} L_\odot $.
Unfortunately, measurement of absolute fluxes for [O I] to compare with this 
prediction 
are not available (Solf 2002, private communication). 
 Inspection of the relative intensities of [O I]
along the slit with PA=0$^{\circ}$ centered
on the star in \cite{B94} indicates
that the region at $\sim $ 2 arcsec south is $\sim$ 4-5 times 
fainter than the star.
We can estimate the [O I] stellar luminosity from the mass
loss rate $\dot {\rm M_w}$, using the relationship 
log $\dot {\rm M_w}=-4.3 + {\rm log}(L_{[O\; I]}/L_\odot)$
\citep{H95}, assuming that outflow rate scales as $\sim$ 0.1 
the accretion rate \citep{C98}, and obtaining the mass accretion
rate from the accretion luminosity. This finally
leads us
to an estimate of $\rm L_{[O I]} \sim 2 \times 10^{-4} L_\odot$ for the region,
which is consistent with our expected limits of the [O I] luminosity.

\section{Discussion}

\subsection{The northern H$_2$ emitting region}

Region N appears to be aligned along the bright rims of the
reflection nebulosity around T Tau N, \S \ref{espacial};
which was plausibly interpreted by S98 as the
walls
of a cavity driven into the surrounding medium by the outflow from
this star.
The H$_2$ emission requires some heating to excite
molecules into the level which can fluoresce with Lyman $\alpha$.  
At distances of 
$\le 50$ AU where the fluorescent emission is observed, local
dust temperatures resulting from heating by the central star(s)
are predicted to be $\sim 100$~K \citep{C94},
whereas excitation temperatures must be $\ge 800$~K by the
argument of the previous section.  
The most likely explanation is that shock heating is responsible
for the excitation of H$_2$ in these regions, as also suggested
by the consistency of the observed spectrum with that of a low-excitation
Herbig-Haro object (\S \ref{espacial}).

 The impact of the wind emanating from T Tau N on the
molecular medium can naturally explain the needed excitation of H$_2$,
and it has already being suggested as a possible explanation
for the near infrared H$_2$ bright rims that seem to
coincide with the scattered light image at
larger scales \citep{H97}.
\cite{B94} 
argued that the E-W outflow is a highly
collimated, high velocity $\sim$ 200 ${\rm km s^{-1}}$
jet at 0.3" $\sim$ 40 AU from the star from analysis of the 
forbidden line emission.
Alternatively, as proposed previously by \cite{S95}, \cite{S98}, and more 
recently by \cite{S02}, the jet results from a density enhancement along the 
outflow
axis in a wide-angle wind. Our observations
support the second hypothesis in that a wide-angle
wind is required to produce the 
excitation along the walls of the cavity. Moreover,
we expect that an oblique shock
would be formed at the interaction of the wide-angle
wind and the molecular environment, as envisoned
by \cite{C80}. Only the normal component of the
velocity would be thermalized,  reducing
the effective pre-shock velocity.
The velocities required to produce
a low excitation spectrum as that emitted along the
cavity (\S \ref{espacial}) are $\le$ 60 km s$^{-1}$
\citep{H87}; higher velocities would result in high
ionization UV lines, such as CIV and Si IV, which are clearly absent in
our data.
Thus, the observed H$_2$ fluorescent emission at the cavity walls 
may be the clearest (though indirect) evidence to date for 
the presence of low-density, 
wide-opening-angle outflows driving 
cavities into the molecular medium in star-forming regions. 

Observations of other shock-excited features close to T Tau N
are needed to help constrain wide-angle wind properties.

\subsection{The southern H$_2$ emitting regions}

The double arcs of region S are located
as close as 0".3 (49 AU) to T Tau S. 
The shape and orientation of this double arc are very similar to the arcs
of H$_2$ 2.12 {\rm $\mu m$} and optical forbidden line emission
described by \cite{H96,H97},
interpreted as
terminal shocks of H$_2$, [S II] and [Fe II] \citep{H97,R95} in the N-S
outflow. 
Figure 4b shows
an enlarged view of the region around the T Tau
system, indicating the positions
of regions N and S and
the shocked structures described by \cite{H97},
as well as the scattered light image of S98.
As mentioned in \S \ref{physical},
the arcs of region S are located inside the
region of forbidden line emission labeled
D by \cite{B94}, very near the centroid.
The outlines of this region as well as the
position of the centroid are indicated
in Figure 4b. 
The mean velocity of region D
is v = -44 km s$^{-1}$,
enough to have a J-shock
which could excite the fluorescence as in
other low excitation HH objects \citep{C95}.
In the case of HH47A, 
\cite{C95} find that 90 \% of the UV line emission comes from a postshock 
distance of $\sim 2 \times 10^{14} {\rm cm} \sim $ 13 AU, 
which we could not resolve. Therefore, the coincidence
of the H$_2$ UV emission with the centroid of the [S II] emission of component
D in \cite{B94} is to be expected.
More sensitive near-infrared observations to constrain the 2.12 $\mu$m 
emission would provide better constraints on column densities in these
regions.

\section{Summary}

 We describe the spatial distribution of the H$_2$
fluorescent emission in the inner 2"x2" region
around the T Tau system. A northern
region of emission coincides with the walls of the
cavity of the envelope seen in scattered light.
Lyman $\alpha$ from T Tau N appears to be strong enough to
excite the fluorescence, but local
temperatures  $\sim$ 1000 K are required to mantain the
fluorescence; shock heating is requred
since stellar heating is not sufficient.
The required
temperatures and the spectrum
of the nebulosity, which resembles that of a low
excitation HH object, indicate a low velocity
shock. The shock may be produced by the interaction
of a wide-angle wind
emanating from T Tau N and the molecular
material around it. This represents the
clearest evidence for the presence of wide-angle
winds at scales $<$ 100 AU from the star.
A southern region of emission consists of
two arcs, with shape and orientation
similar to the arcs of H$_2$ 2.12 {\rm $\mu m$}
and forbidden line emission crossing
the outflow associated with the T Tau S system.
The arcs are near the centroid of the
forbidden line emission in the
blueshifted lobe of the outflow. The 
velocity of the outflow around the arcs
are consistent with J-shocks, which
could power the fluorescence as in
low excitation Herbig-Haro objects.

\vskip 0.5in

\noindent
Acknowledgments. 
We wish to thank K. Stapelfeldt for providing us the optical image of T 
Tau, M. Kasper for the infrared images and spectra of T Tau N, 
and K. B\"ohm and J. Solf for the original 
long slit spectrograms of T Tau N and its surroundings.  We also thank H. 
Abgrall, for the updated tables 
of Einstein coefficients and wavelengths for the UV H$_2$ transitions. 
This work was supported
by NASA through grants GO-08206.01-97A and GO-08317.01-97A from the
Space Telescope Science Institute, and by NASA Origins of Solar Systems
grant NAG5-9670.
The non HST ultraviolet data used in this paper were obtained from the
Multimission Archive at the Space Telescope Science Institute (MAST).
STScI is operated by the Association of Universities for Research in
Astronomy, Inc., under NASA contract NAS5-26555. Support for MAST for
non-HST data is provided by the NASA Office of Space Science via grant
NAG5-7584, and by other grants and contracts.
  

\clearpage
\begin{deluxetable}{cccccccc}
\tabletypesize{\scriptsize}
\tablecolumns{8}
\tablewidth{0pt}
\tablecaption{Integrated Observed UV
Fluxes coming from de-excitation of (1-2)P(5)}
\tablehead{
\colhead{ y offset }&  \multicolumn{6}{c}{{\rm Flux} ($\times 10^{-15}$ 
{\rm 
erg s$^{-1}$ 
cm$^{-2}$ }) {\rm by 
transition} }& \colhead{}}
\startdata
 "  & (1-6)R(3) & (1-6)P(5)  & (1-7)R(3) & (1-7)P(5) & (1-8)R(3) & 
(1-8)P(5) &  
$\frac{F_{(1-6)P(5)}}{F_{(1-7)P(5)}}\tablenotemark{a}$\\ 
  & 1431.01 {\AA} & 1446.12 {\AA}  & 1489.56 {\AA} & 1504.75 {\AA} & 
  1547.33 {\AA} & 1562.39 {\AA}  & \\
\cline{1-8}
0.39   & 0.4 & 1.4 & 1.1 & 2.5 & 1.0 & 2.0 & 0.75\\
-0.39  & 0.2 & 1.0 & 1.1 & 1.8 & 1.2 & 0.6 & 0.81\\
-0.63  & 0.5 & 2.0 & 0.6 & 3.4 & 1.5 & 2.6 & 0.78\\ 
-0.86  & 1.2 & 0.6 & 1.1 & 1.3 & 1.6 & 1.5 & 0.74\\ 
-1.10  & 0.5 & 1.0 & 1.7 & 1.8 & 1.2 & 0.5 & 0.78\\ 
-1.33  & 0.4 & 1.4 & 0.2 & 2.6 & 0.7 & 0.4 & 0.75\\ 
-1.57  & 0.6 & 0.4 & 0.6 & 0.8 & 0.3 & 0.7 & 0.85\\
\enddata 
\tablenotetext{a}{Assuming a de-reddening correction with $A_v$=2.2} 
\end{deluxetable} 


\clearpage
\begin{figure}
\plotone{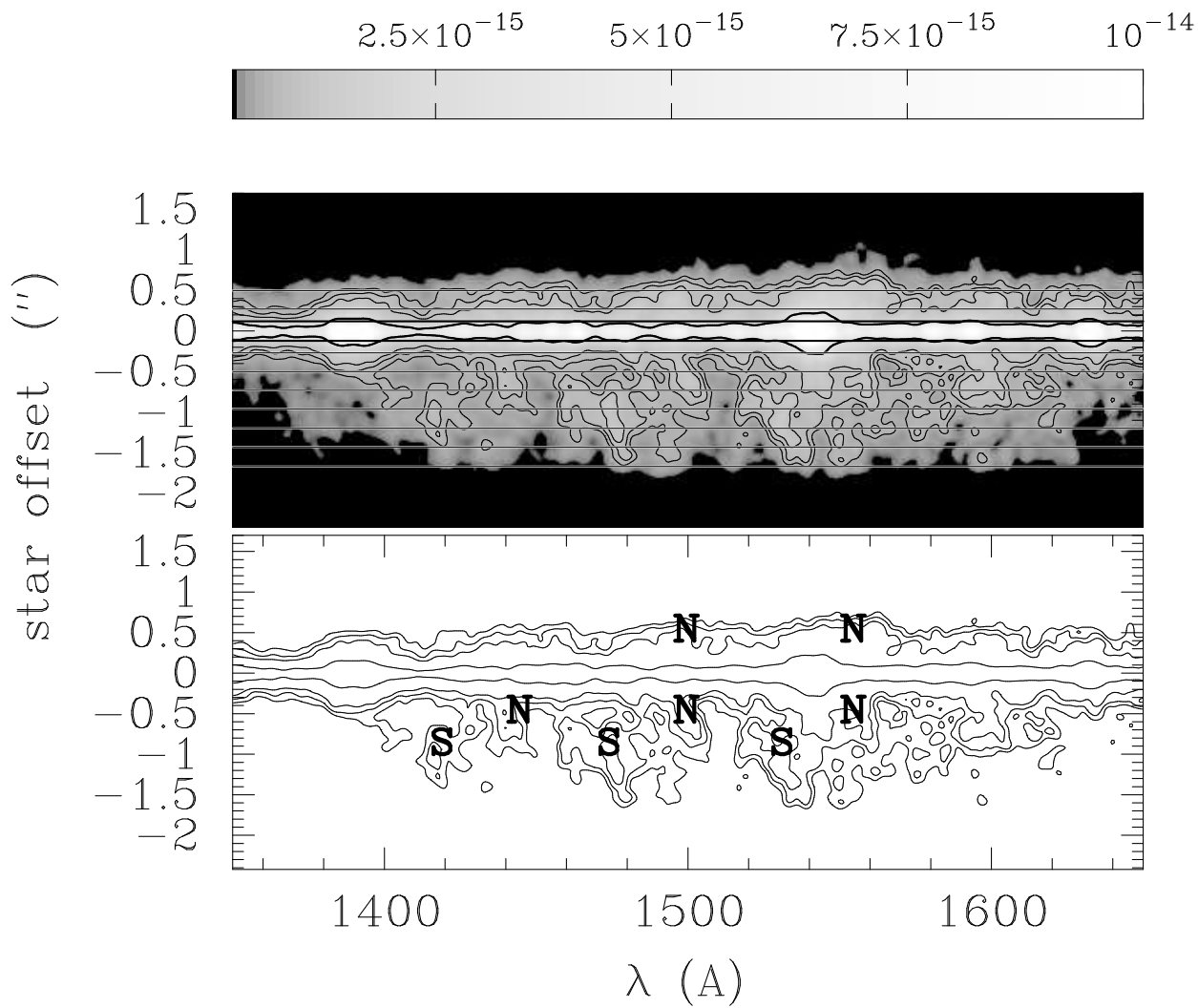}
\figcaption{\scriptsize Position-dispersion image of T Tau N
between 
1350 and 1650 \AA.  
The  upper panel shows the
observed spectrum. The isocontours have values of 0.65e-15, 0.8e-15 1.0e-15 and
3.0e-15 erg cm$^{-2}$ s$^{-1}$ \AA$^{-1}$ pix$^{-1}$.
The extended
emission is evident,
specially at negative offsets from the star.
The lower panel 
shows the same isocontours,
identifying some the features described in the text.
Note the groups of periodic features, one close 
to the stellar spectrum, marked N, and 
another to the south, marked S,
displaced redwards respect to the first. 
The location of the extraction boxes used to obtain the envelope
spectra in indicated in the upper panel.}
\end{figure}

\clearpage
\begin{figure}
\plotone{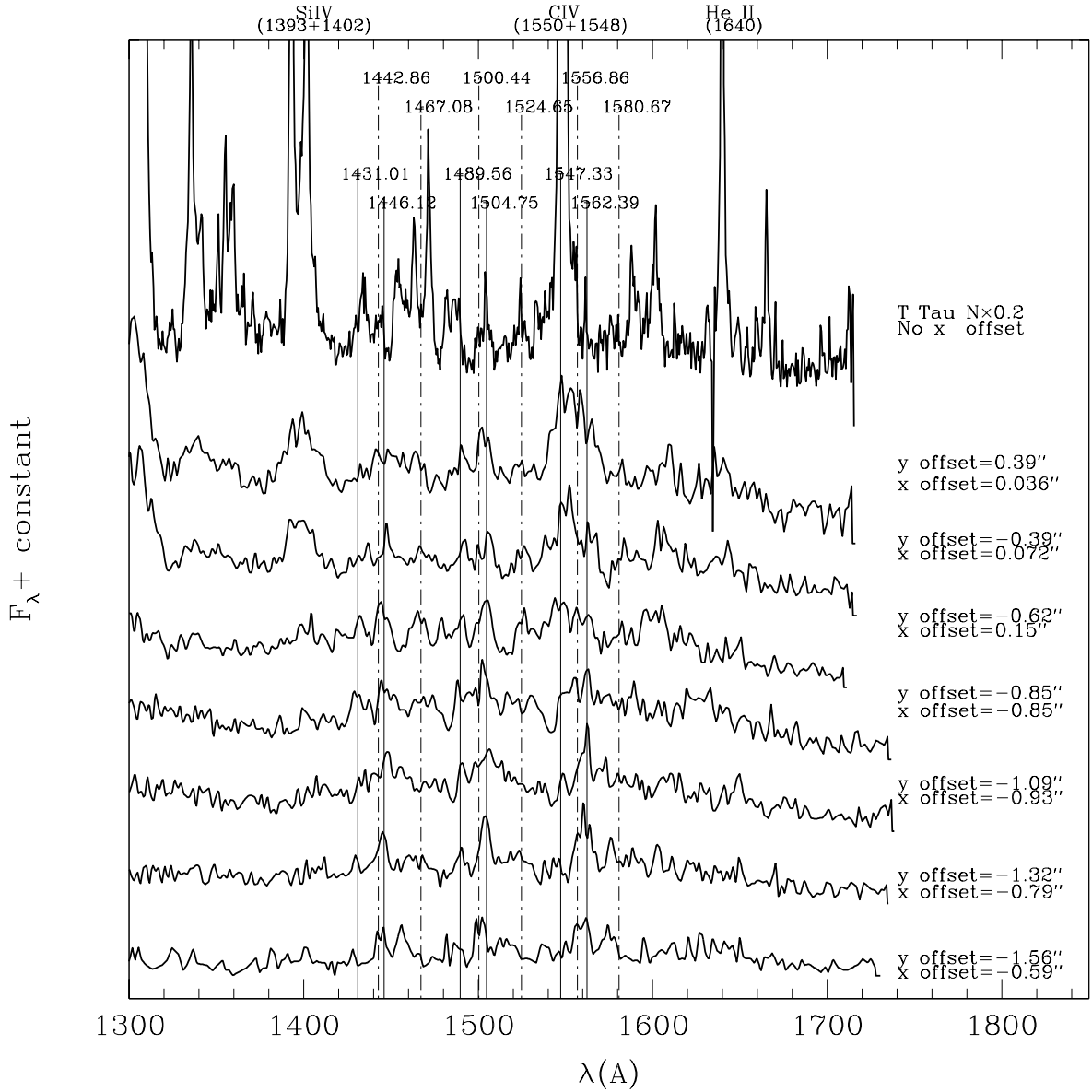}
\figcaption{\scriptsize Spectra of T Tau N and of the 
extended emission at 
several offsets from the star, indicated in Figure 1. The y offset is the 
position along the 
slit where the spectrum 
was extracted. 
Lines of Si IV, C IV and He II can be identified in the spectrum of T Tau 
N, 
as well as in 
offsets {\it y}=+0.39", -0.39" and -0.62"; the emission from these ionized 
elements 
in the nebulosity is due to the stellar PSF wings. 
Several H$_2$ fluorescent lines are present in 
the stellar spectrum and in 
the spectra from the nebulosity, with a much higher flux
than expected from the stellar PSF, indicating local emission.
The solid lines indicate fluorescent
lines arising from the (1-2)P(5) transition,
while the dot-dashed lines indicate lines from the (1-2)R(6)
transition.
The spectra of the offsets need to be shifted in order 
to make the identifications, resulting in a spatial shift given by {\rm x} 
offset (see \S 
2). It 
can be seen that there are two different loci for the emission, one at 
positive  and the 
other at negative offsets.
All offsets are measured relative to the
position of T Tau N.}
\end{figure}
\clearpage
\begin{figure}
\plotone{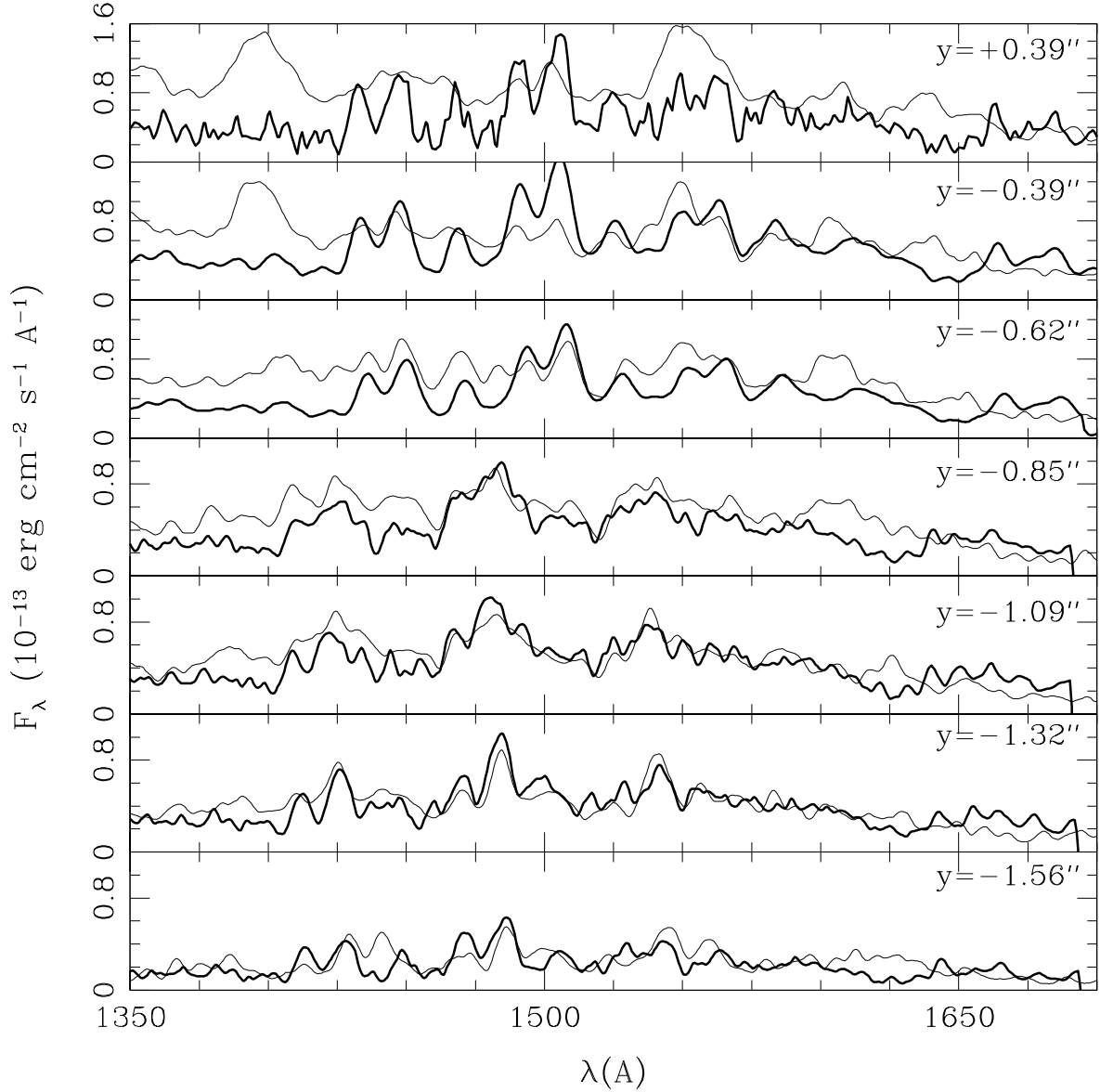}
\figcaption{\scriptsize De-reddened spectra 
and model predictions for 
offsets along the slit south of T Tau N, following
the extended nebulosity.  The thin
line corresponds to the observations, and the thick
line to the best fit for the simulated emission, using the spectrum
of HH 43 as template.  The first three boxes show the contribution of the 
fluorescence 
following
the optical envelope (region N in Figure 4). The emission at the wavelengths of 
CIV 1548+1550 
{\AA}, SiIV 
1396+1402 {\AA}, and HeII 1640 {\AA} correspond to the stellar PSF. The last 
four boxes 
show the contribution of the
double arc-like structure (region S in Figure 4). }
\end{figure}

\clearpage
\begin{figure} 
\plottwo{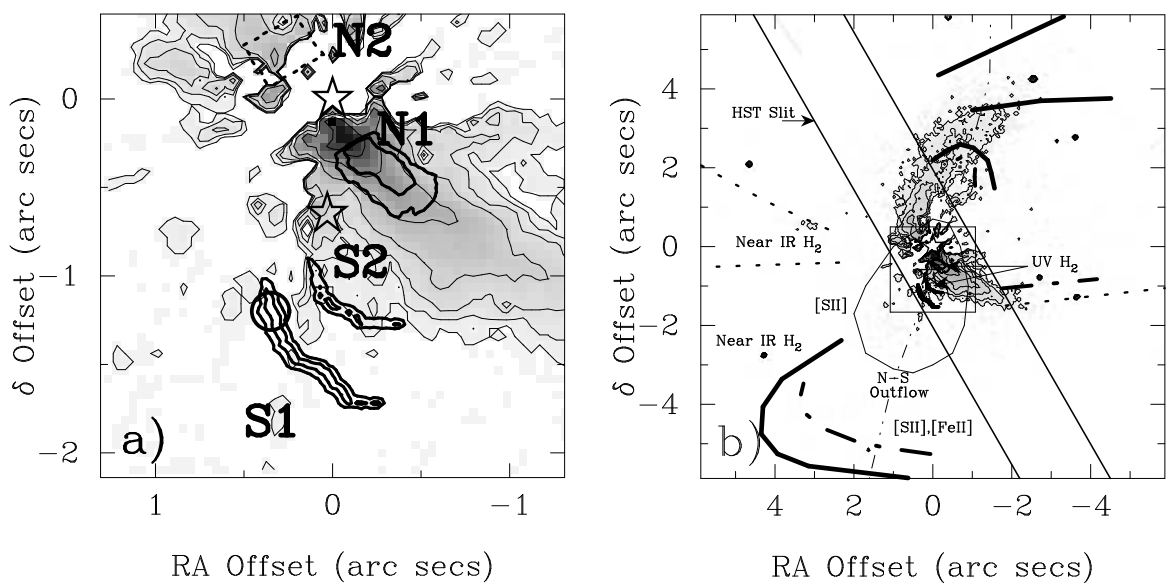}{f4b.ps}
\figcaption{\scriptsize Spatial distribution of the H$_2$ 
emission.
a): Isocontours of
fluorescent H$_2$ emission (thick solid lines) superimposed on
magnitude isocontours of
0.55 $\mu$m
emission from Stapelfeldt et al. (1998) (relative intensity grayscale/thin 
lines).
The isocontours
for the fluorescent features show relative intensity, with steps of 1 
magnitude, corresponding to values of 1, 0.398 and 0.158.
The regions discussed in the text, N1, N2, S1, and S2 
are indicated.
The mean centroid
position of component D from B\"ohm \& Solf (1994) is also
shown (open circle).
The dotted box  shows an estimate for the location of region N2, since the 
overextraction 
of 
the star in the optical image makes it difficult to obtain a brightness 
distribution (S98). 
b): View of a larger area around the region shown
in the upper panel (indicated by the box), showing
the observing slit and features discussed in the literature:
Terminal H$_2$ shocks (thick lines), [S II] or [Fe II] shocks (dot-dashed 
line),
oblique H$_2$  shocks (dashed lines), from Herbst et al. (1997),
associated to the N-S outflow.
The extend of of component D from B\"ohm \& Solf (1994)
is also shown (lobe shaped contour).}
\end{figure}

\clearpage
\begin{figure}
\plotone{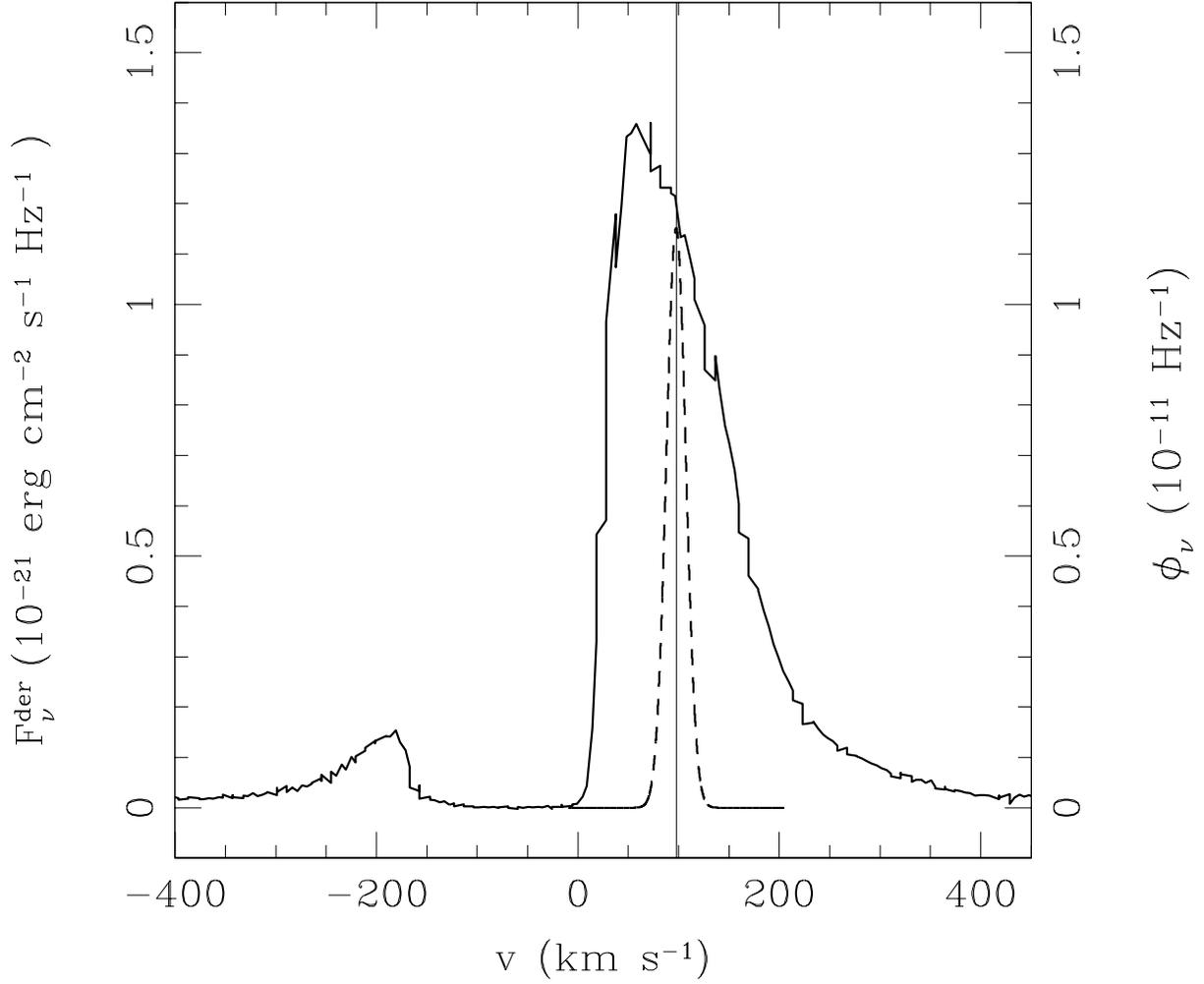}
\figcaption{\scriptsize De-reddened high resolution 
line profile of Mg II k $\lambda$ 2796.3 in T
Tau N,
obtained with STIS/NUV-MAMA and echelle grating E230M
in $HST$ program GO8627 (solid heavy line).
The 
vertical light solid line indicates
the central velocity of the 
radiation pumping transition (1-2) P(5).
The H$_2$ line profile $\phi_\nu$ is indicated
(dashed line).
}
\end{figure}


\clearpage
\begin{thebibliography}{}

\bibitem[Abgrall et al. (2000)]{A00}
 Abgrall, H.,
Roueff, E., \& Drira, I., 2000 A \& A.  Sup.  141, 297 
\bibitem[Ardila et al. (2002)]{A02}
Ardila, D., Basri, G., Walter, F., Valenti, J., \&Johns-Krull, C., 2002, 
ApJ, 
566, 1100


\bibitem[B\"ohm \& Solf (1994)]{B94}
B\"ohm, K., \& Solf., J., 1994, ApJ 430, 277-290

\bibitem[Brown et al. (1981)]{B81}
 Brown, A., Jordan, C, Millar, T.  J., \& Gondhalekar, P.,
Wilson, R., 1981, Nature, 290,34 

\bibitem[Calvet et al. (1994)]{C94}
 Calvet, N., Hartmann, L., Kenyon, S.  J.,
\& Whitney, B.  A., 1994, ApJ 434,330  

\bibitem[Calvet  (1998)]{C98} Calvet, N., 1998, Eight Astroph. Conf., Accretion 
Processes in Astrophysical Systems: Some Like it Hot!, 431, 495-504

\bibitem[Calvet et al. (2003)]{C03}Calvet, N. et al., in preparation.

\bibitem[Cant\'o (1980)]{C80} Cant\'o, J.\ 1980, A \& A, 86, 327 


\bibitem[Curiel et al. (1995)]{C95}
 Curiel, S., Raymond, J., Wolfire, M., Hartigan, P.,
Schwartz, R., \& Nosenson, P., 1995, ApJ, 453, 322-331 


\bibitem[Dyck et al. (1982)]{D82}
Dyck, H., Simon, T., Zuckerman, B., 1982, ApJ 255, L103-106

\bibitem[Duch\^ene et al. (2002)]{D02}
Duch\^ene, G., Ghez, A., \& McCabe, C., 2002, ApJ 568, 771

\bibitem[Eisl\"offel \& Mundt (1998)]{E98}
Eisl\"offel, J., \& Mundt,
R., 1998, AJ 115, 1554 

\bibitem[Gautier et al. (1976)]{G76}
Gautier III, T.  N., Fink, U., Treffers, R., 
\& Larson, H.P., 1976, ApJ 207, L219-L133 

\bibitem[Gullbring et al. (2000)]{G00}
Gullbring, E., Calvet, N., Muzerolle, J., Hartmann, L., 2000, ApJ 544, 
927

\bibitem[Hartigan et al. (1987)]{H87}
Hartigan, P., Raymond, J., \&
Hartmann, L., 1987, ApJ, 316, 323 

\bibitem[Hartigan et al. (1995)]{H95}
Hartigan, P., Edwards, S., \&
Ghandour, L., 1995, ApJ, 452, 736 

\bibitem[Herbst et al. (1996)]{H96}
Herbst, T., Beckwith, S., \& Krabbe, A.,
1996, AJ 111,2403 

\bibitem[Herbst et al. (1997)]{H97} 
Herbst, T., Robberto, M., \& Beckwith, 1997 AJ 114, 744

\bibitem [Herzberg (1950)]{H50}
Herzberg, G., 1950, in \emph{Spectra of Diatomic Molecules}, Second
Edition, Ed.  Van Nostrand Reinhold Company 

\bibitem[Jordan et al. (1978)]{J78}
 Jordan, C., Brueckner, G. E., Bartoe, J.-D.  F., Sandlin, G.  D., \& van 
Hoosier, 
M.  E.,
1978 ApJ 226, 687-697 

\bibitem[Kasper et al. (2001)]{K01}
 Kasper, M. E., Feldt, M.,
Herbst, T.  M., \& Hippler, S., 2001, astro.ph.12042K, to appear in ApJ, 
March
2002 

\bibitem[Kenyon et al. (1994)]{K94}
Kenyon, S.  J., Dobrzycka, D., \& Hartmann, L., 1994, AJ, 108,1872 

\bibitem[K\"ohler et al. (2000)]{Kh00}
K\"ohler, R., Kasper, M., \& Herbst, T., 2000, Poster Proceedings of IAU Symp. 
200 on The Formation of Binary Stars, eds. B. Reipurth \& H. Zinnecker, p. 63

\bibitem[Koresko (2000)]{K00}
Koresko, C., 2000, ApJ 531, L147

\bibitem[Li and Shu (1996)]{L96}
Li, Z.-Y., \& Shu, F., 1996, ApJ, 468, 261-268

\bibitem[Momose et al. (1996)]{M96}
Momose, M., Nagayoshi, O., Kawabe, R., Masahiko, H., \& Nakano, T., 1996, 
ApJ, 
470, 1001-1014

\bibitem[Osorio et al. (2002)]{O02}
Osorio, M., D'Alessio, P., Calvet, N., \& Hartmann, L., 2002, ApJ, 
submitted

\bibitem[Reipurth et al. (1997)]{R97} Reipurth, 
B., Bally, J., \& Devine, D.\ 1997, AJ, 114, 2708 

\bibitem[Robberto et al. (1995)]{R95}
Robberto, M., Clampin, M., Ligori, S., Paresce, F., Sacc\'a, V., \&
Staude, 1995 A \& A, 296, 431 

\bibitem[Schwartz (1983)]{Sc83}
Schwartz, R., 1983, RMAA, 7,27-54 

\bibitem[Shang et al. (1998)]{S98}
Shang, H., Shu, F., \& Glassgold, A., 1998, ApJ, 493, L91-94


\bibitem[Shang et al. (2002)]{S02}
Shang, H., Glassgold, A., Shu, F., \& Lizano, S., 2002, ApJ 564, 853

\bibitem[Shu et al. (1995)]{S95} Shu, 
F. H., Najita, J., Ostriker, E.~C., \& Shang, H.\, 1995, ApJ, 455, L155 

\bibitem[Solf, B\"ohm, \& Raga(1988)]{B88} Solf, J., Bohm, 
K. H., \& Raga, A., 1988, ApJ, 334, 229 

\bibitem [Solf \& B\"ohm (1999)]{S99}
Solf, J., \& B\"ohm, K.  H., 1999, ApJ
523,709 

\bibitem[Stapelfeldt et al. (1998)]{St98}
Stapelfeldt,
K., Burrows, C., Krist, J., Watson, A., Ballester, G., Clarke, J., 
Crisp., D.,
Evans, R., Gallagher, J., Griffiths, R., Hester, J., Hoessel, J., 
Holtzman, J.,
Mould, J., Scowen, P., Trauger, J., \& Westpal, J., 1998, ApJ 508, 74 

\bibitem[Valenti et al. (2000)]{V00} 
Valenti, J., Johns-Krull, C.M., \& Linsky, J., 2000, ApJS, 129,399 

\bibitem[Whitney \& Hartmann (1993)]{W93}
Whitney, B.  A., \& Hartmann, L., 1993, ApJ 402, 605-622 

\bibitem[Wood, Karovska, \& Raymond (2002) ]{W02}
 Wood, B.~E., Karovska, M., \& Raymond, J.~C.\ 2002, ApJ, 575, 1057

\end{thebibliography}
\end{document}